\input{aipcheck}

\documentclass[
    ,final            
  ]
  {aipproc}

\layoutstyle{6x9}
\newcommand{\1}{{\rm 1\hspace*{-0.4ex}%
\rule{0.1ex}{1.52ex}\hspace*{0.2ex}}}

\begin{document}

\title{Sigma meson and lowest possible glueball candidate in an extended 
linear $\sigma$ model}


\author{Tamal K. Mukherjee}{
  address={Institute of High Energy Physics, Chinese Academy of 
Sciences, Beijing, China}, altaddress={Theoretical Physics Center 
for Science Facilities, Chinese Academy of Sciences, Beijing, China}
}

\author{Mei Huang}{
  address={Institute of High Energy Physics, Chinese Academy of 
Sciences, Beijing, China}, altaddress={Theoretical Physics Center 
for Science Facilities, Chinese Academy of Sciences, Beijing, China}
}

\author{Qi-Shu Yan}{
  address={College of Physical Sciences, Graduate University of Chinese 
Academy of Sciences, Beijing, China}
}

\begin{abstract}
We formulate an extended linear $\sigma$ model of a quarkonia nonet and a 
tetraquark nonet as well as a complex iso-singlet (glueball) field to study 
the low-lying scalar meson. Chiral symmetry and $U_A(1)$ symmetry and their 
breaking play important role to shape the scalar meson spectrum in our work. 
Based on our study we will comment on what may be the mass of the lowest 
possible scalar and pseudoscalar glueball states. We will also discuss on 
what may be the nature of the sigma or $f_0(600)$ meson.
\end{abstract}

\maketitle


\section{Introduction}

Theoretical understanding of the so-called Higgs Boson of QCD namely 
the $\sigma$ or $f_0(600)$ meson is important as they play an important 
role in chiral symmetry breaking and can be used as a 
probe for QCD vacuum. While the confirmation of its existence from 
$\pi \pi$ scattering process puts an end to the decades-long controversy 
\cite{Caprini:2005zr,Scalar}, the agreement on the nature of 
$\sigma/f_0(600)$ however has not yet been achieved. The tetraquark model 
\cite{jaffe} supports $\sigma/f_0(600)$ to be a tetraquark state. 
Whereas, recent data from $\pi \pi$ and $\gamma \gamma$ scattering 
\cite{Minkowski:1998mf,arXiv:0804.4452} suggest it has a sizable 
fraction of glueball and another study from  K-matrix analysis 
\cite{Mennessier:2010xg} suggested that $f_0(600)/\sigma$ should be 
a glueball dominant state. 

On the other hand the agreement on the lowest possible iso-scalar and 
pseudoscalar glueball states has neither been acheived. The result from 
Lattice simulations suggests $f_0(1500)$ or $f_0(1700)$ could be a 
glueball rich iso-scalar, while $\eta(1489)$ can be a glueball rich 
pseudoscalar \cite{Mathieu:2008me}.

The great success of chiral symmetry in understanding the nature of
lightest pseudo-scalars motivates us to use an extended version of 
linear $\sigma$ models \cite{qishu,Fariborz:2011es} to study the nature 
of these low-lying isoscalars and pseudoscalars. The motivation for 
constructing an extended linear sigma model consisting of effective 
quarkonia, tetraquark and glueball fields comes from the physical 
considerations that scalar condensates are allowed by the QCD vacuum. 
So in principle apart from quarkonia and tetraquark condensates, scalar 
glueball condensate should also be present in the QCD vacuum and need to 
be considered while studying the vacuum excitation.
The mixing pattern followed in our framework 
comes from the consideration that mesons having identical external 
quantum numbers can mix even if they have different internal flavour 
structures. Based on these premises we attempt to study the nature of 
$\sigma/f_0(600)$ meson as well as lowest iso-scalar and pseudoscalar 
glueball candidate. In the following section we briefly review the model 
Lagrangian and the symmetry breaking pattern. At last we will 
present our results and conclude.

\section{Model}
The extended linear $\sigma$ model can be systematically formulated under 
the symmetry group $SU_R(3) \times SU_L(3) \times U_A(1)$.
Three types of chiral fields are included: a $3\times3$
matrix field $\Phi$ which denotes the quarkonia states, a $3\times3$
matrix field $\Phi^\prime$ which denotes the tetraquark state, and a
complex chiral singlet field $Y$ which denotes the pure glueball states.

Up to the mass dimension $O(p^4)$ (we assume that they are the most 
important operators to determine the nature of light scalars of ground 
states), the Lagrangian of our model can include two parts: the symmetry 
invariant part ${\cal L}_S$ and the symmetry breaking one ${\cal L}_{SB}$.
The symmetry invariant part includes those terms which respect 
$SU(3)_L \times SU(3)_R$ symmetry as well as $U_A(1)$ symmetry:

\begin{eqnarray}
\label{sl} 
{\cal L}_{S} &= Tr(\partial_\mu \Phi \partial^{\mu} \Phi^{\dagger}) 
+ Tr(\partial_\mu \Phi^\prime \partial^{\mu} \Phi^{\dagger \prime})   
+ \partial_\mu Y \partial^\mu Y^\star
- {m_\Phi}^2 Tr (\Phi^\dagger \Phi) 
- {m_{\Phi^\prime}}^2 Tr (\Phi^{\dagger \prime} \Phi^\prime) \nonumber \\
&-{m_Y}^2 Y Y^\star 
-\lambda_1 Tr (\Phi^\dagger \Phi \Phi^\dagger \Phi)
-{\lambda_1}^\prime Tr (\Phi^{\dagger \prime} \Phi^\prime \Phi^{\dagger \prime} \Phi^\prime)
-\lambda_2 Tr (\Phi^\dagger \Phi \Phi^{\dagger \prime} \Phi^\prime)
- \lambda_Y (Y Y^\star)^2 \nonumber \\
& - [ \lambda_3 \epsilon_{abc} \epsilon^{def} {\Phi_d}^a {\Phi_e}^b
{\Phi_f}^{\prime c} + h.c.] + [ k \,\, Y Det(\Phi) + h.c.],
\end{eqnarray}

While the symmetry breaking part includes the following terms
\vskip -0.2in
\begin{equation}
{\cal L}_{SB} = [Tr(B.\Phi) + h.c.] + [Tr(B^\prime.\Phi^\prime) +
h.c.]  + (D.Y +h.c.) - [ \lambda_m Tr(\Phi \Phi^{\dagger \prime})
+h.c. ].
\label{sbt}
\end{equation}
To construct the symmetric terms in the Lagrangian we closely follow 
\cite{fariborz1} where the choice of terms are limited by the number 
of internal quark plus antiquark lines at the effective vertex, which 
is set to 8. This condition is relaxed for two terms with coupling 
constants ${\lambda_1}^\prime$ and $\lambda_2$ and the reason to 
include these two terms stems from the practical consideration of 
making sure our potential is bounded. For detailed discussion on the 
terms appearing in the Lagrangian please refer to \cite{OurPaper}. 
As evident from the Lagrangian both chiral symmetry $SU(3)_L \times 
SU(3)_R $ and $U(1)_A$ symmetry are explicitly broken by the terms 
in ${\cal L}_{SB}$. The $3 \times 3$ matrices $B$ 
and $B^\prime$ responsible for the breaking of the symmetry can be 
parametrized as: $B (B^\prime) = T_a b_a (T_a {b_a}^\prime)$ ($T_a$ 
are the generators of $U(3)$ with $T_0 = \sqrt{\frac{1}{6}} 
 \1$). 

Since the vacuum expectation values of the quarkonia and tetraquark fields 
can carry those quantum numbers which are allowed by the QCD vacuum, only 
$a = (0,3,8)$ fields are allowed. The choice of this external fields 
$b_a ({b_a}^\prime)$ control the nature and extent of the symmetry breaking. 
Out of various possible symmetry breaking scenarios, we consider the following 
case in our present study where:

\begin{itemize}
\item $b_0 ({b_0}^\prime) \neq 0$,  $b_3 ({b_3}^\prime) = 0$ and 
$b_8 ({b_8}^\prime) \neq 0$. In this case, $U(3)_A$ and $SU(3)_V$ both are 
explicitly broken and $SU(3)_V$ is broken to $SU(2)_V$. As a result 
$m_u = m_d \neq m_s$, where $m_i$ is the quark mass of the $i^{th}$ flavour.
\end{itemize}

This is reasonable considering the up and down quark masses are nearly equal 
to each other and thereby indicating $SU(2)_V$ is a good (approximate) symmetry.
The remnant $SU(2)_V$ isospin symmetry allows us to represent
two condensates for quarkonia and teraquark fields each as: $v_0$,
$v_8$ and ${v^\prime}_0$, ${v^\prime}_8$ respectively. While the gluonic
condensate in our theory is labeled as $v_y$.

\section{Results and Conclusion}
Due to the unbroken $SU_V(2)$ isospin symmetry, physical scalar and
pseudo-scalar states can be categorized into three groups with
isospin quantum numbers as $I = 1$ (triplet), $\frac{1}{2}$
(doublet) and $0$, respectively. Only bare quarkonia, tetraquark and
glueball fields with the same isospin quantum number can mix with
each other to form physical states. Moreover, there is no mixing
between scalar and pseudoscalar fields. Thus the chiral singlet
glueball field can only mix with the isospin singlets of quarkonia
and tetraquark fields. Using these facts, the physical states below 
2 GeV can be tabulated as given in Table \ref{mixing}, where the 
isodoublet $\{K, K^\prime\}$ is connected with the isodoublet 
\{$K^*,K^{*\prime} $\} by charge conjugation. And a similar relation holds
for $\{\kappa$, $\kappa^\prime\}$ and $\{\kappa^*$, $\kappa^{*\prime}\}$.

\begin{table}
\begin{tabular}{ |c|c|c|c| }
\hline
Isospin & $I=1$ & $I=\frac{1}{2}$ & $I=0$  \\ \hline
PseudoScalars(P=-1)& $\{\pi, \pi^\prime \}$ & \{K, $K^\prime$\}, 
\{$K^*, K^{*\prime} $\} & \{$\eta_1, \eta_2, \eta_3,
\eta_4, \eta_5$ \} \\ \hline
Scalars(P=1)& \{$a$, $a^\prime$ \}, & \{$\kappa$, $\kappa^\prime$\}, 
\{$\kappa^*$, $\kappa^{*\prime}$\}, & $\{f_1,f_2,f_3,f_4,f_5\}$ \\ \hline
\end{tabular}
\caption{The categorization of scalar and pseudo-scalar states in term of 
isospin quantum number are demonstrated. States in the same category can 
mix with each other.}
\label{mixing}
\end{table}

There are 15 parameters in our model. To solve these parameters we treat 
the tetraquark vacuum condensates as well as the mixing angles for 
isotriplet ($\theta_{\pi}$) and isodoublet ($\theta_K$) fields as input 
parameters. Then from the mass matrices of $a, a^\prime$, $\kappa$ and 
$\kappa^\prime$, parameters related to isotriplet and isodublet sectors 
are solved. The symmetry parameters $\{b_0,b_8,b_0^\prime,b_8^\prime, D\}$ 
are solved from the vacuum stability conditions. Whereas, to solve for 
the parameters related to glueball sector, we use following two conditions: 

\begin{equation}
Tr[{M_\eta}^2]_{Model} = Tr[{M_\eta}^2]_{Exp}\,\,, \label{trace} 
\hskip 0.5in
Det[{M_\eta}^2]_{Model} = Det[{M_\eta}^2]_{Exp}\,\,.
\end{equation}
At the end we are left with one free parameter which is bare glueball 
mass $m_Y$ and is used in our study as a scanning parameter.

For space constriant we refer to \cite{OurPaper} for details on 
parameter fixing and method to choose the best fit solution. In choosing 
the best fit solution we vary the $\pi^\prime$ mass from $1.2-1.4$ GeV 
and choose those solution which give the tree level decay width for 
$f_0(600)\rightarrow \pi \pi$ between $0.35-0.9$ GeV.

Our best fit parameter set is presented in Table (\ref{chi1-prm}).
We would like to highlight a few features out from it. 1) In absence 
of the explicit symmetry breaking terms, it is the negative mass 
parameter ${m_\Phi}^2$ that would trigger the spontaneous chiral 
symmetry breaking. 2) The sign of $v_Y$ is correlated with the sign 
of $k$, and the sign of $k$ is determined from the mass spectra of 
pseudoscalar sector. 3) The couplings $\lambda_1$, $\lambda_1^\prime$, 
$\lambda_Y$ are positive which guarantee the potential is bounded from 
below. 4) The values of $\lambda_1$, $\lambda_2$, and $\lambda_Y$ as 
well as $k$ are large, which demonstrate the non-perturbative nature 
of the model. 

\begin{figure}[t]
\centerline{
  \includegraphics[height=5 cm, width = 7 cm]{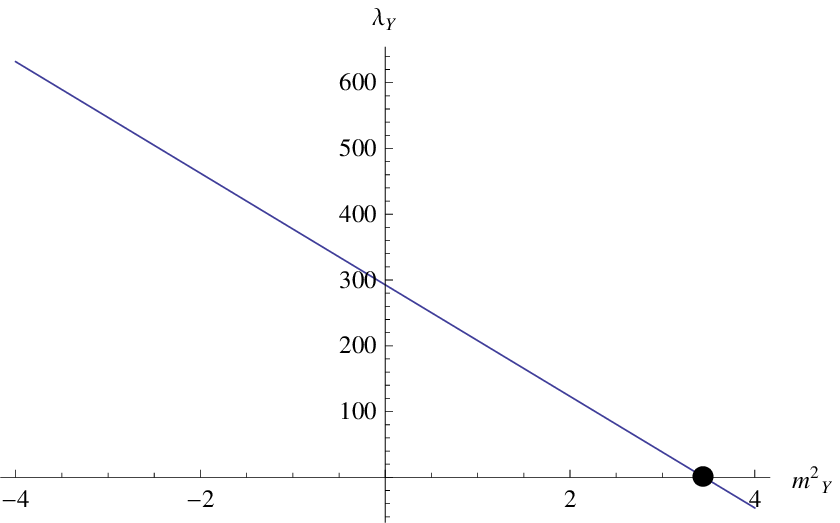}
\hspace*{0.2cm}
  \includegraphics[height=5 cm, width = 7 cm]{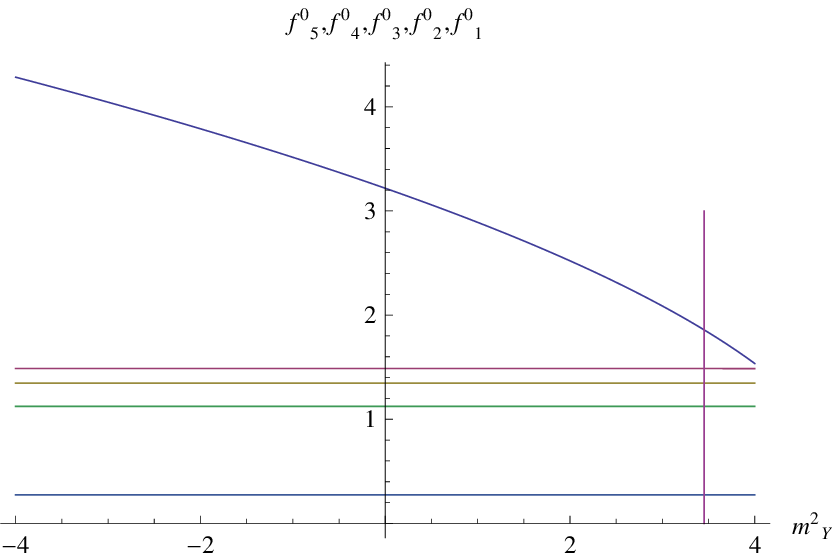}} 
\caption{(Left figure) The dependence of $\lambda_Y$ upon $m_Y^2$ is demonstrated. 
A solid circle marker shows the point $\lambda_Y=0$, which corresponds to 
$m_Y^2=3.452$ ($m_Y=1.858$) . (Right figure) The dependence of mass of $f_0$ upon 
$m_Y^2$ is demonstrated. A vertical line with $m_Y^2$ is drawn to read out the 
lowest mass $m_{f_0^5}=1.86$.}
\label{figf}
\end{figure}

\begin{table} 
\begin{tabular}{ c c c c c c }
\hline
\tablehead{1}{r}{b}{Parameter} & \tablehead{1}{r}{b}{Value} & \tablehead{1}{r}{b}{Parameter} 
& \tablehead{1}{r}{b}{Value} & \tablehead{1}{r}{b}{Parameter} & \tablehead{1}{r}{b}{Value} \\ 
\hline
$\theta_{\pi}$ (radian)          & -0.604   & ${\lambda_1}^\prime$     & 8.248    & ${m_{\Phi}}^2$ ($GeV^2$)         & -0.025   \\ 
$\theta_{K}$ (radian)    	 & -0.714   & $\lambda_2$              & 76.428   & ${b_0}^\prime$ ($GeV^3$)         & 0.166    \\ 
$v_0$ (GeV)             	 & 0.074    & $\lambda_3$ (GeV)        & -0.738   & ${m_{\Phi^\prime}}^2$ ($GeV^2$)  & 0.744    \\ 
$v_8$ (GeV)             	 & -0.115   & $\lambda_Y$              & 38.327   & ${b_8}^\prime$ ($GeV^3$)         & 0.18     \\ 
${v_0}^{\prime}$ (GeV)  	 & 0.203    & k                        & -78.15   & $\lambda_1$                      & 35.465   \\ 
${v_8}^{\prime}$ (GeV)   	 & 0.126    & $\lambda_m$ ($GeV^2$)    & -1.044   & D ($GeV^3$)                      & -0.265   \\ 
$v_y$ (GeV)             	 & -0.109   & $b_0$ ($GeV^3$)          & -0.085    \\ 
${m_Y}^2$ ($GeV^2$)     	 & 3.0      & $b_8$ ($GeV^3$)          & -0.161   \\ \hline
\end{tabular}
\caption{The values of parameters in our fit are shown where the best value of $m_{\pi{^\prime}}$ is found to be $m_{\pi^\prime} = 1.2$ GeV.}
\label{chi1-prm}
\end{table}

\begin{table}  
\begin{tabular}{ c c c c c c }
\hline
\tablehead{1}{r}{b}{Meson} 
& \tablehead{1}{r}{b}{Our Value \\(GeV)}  
& \tablehead{1}{r}{b}{quarkonia ($\%$)} 
& \tablehead{1}{r}{b}{tetraquark ($\%$)} 
& \tablehead{1}{r}{b}{glueball ($\%$)}  
& \tablehead{1}{r}{b}{Experimental \\ Value (GeV)}
 \\ \hline
$\eta_5$ & 1.858  & 0.037 & 0.001 & 99.962 & 1.756 $\pm$ 0.009 \\  
$\eta_4$ & 1.380  & 75.803 & 24.167 & 0.03  & 1.476 $\pm$ 0.004 \\ 
$\eta_3$ & 1.291  & 26.700 & 73.294 & 0.006 & 1.294 $\pm$ 0.004 \\ 
$\eta_2$ & 0.907   & 15.852 & 84.145 & 0.003  & 0.958 $\pm$ $24 \times 10^{-5}$ \\ 
$\eta_1$ & 0.595   & 81.607 & 18.393 & 0.0 & 0.548 $\pm$ $24 \times 10^{-6}$ \\ 
${f_5}^0$ & 2.09   & 0.01 & 0.0 & 99.99  & - \\  
${f_4}^0$ & 1.487   & 77.469 & 22.53 & 0.001 & 1.505 $\pm$ 0.006 \\ 
${f_3}^0$ & 1.347  & 22.177 & 77.82 & 0.003 & 1.2-1.5 \\ 
${f_2}^0$  & 1.124  & 21.561 & 78.439 & 0.0  & 0.980 $\pm$ 0.010 \\ 
${f_1}^0$  & 0.274  & 78.784 & 21.211 & 0.005 & 0.4-1.2 \\ \hline
\end{tabular}
\caption{Mass spectra and components for the pseudo-scalar amd scalar mesons based on our 
fit are shown where the best value of $m_{\pi{^\prime}}$ is found to be 
$m_{\pi^\prime} = 1.2$ GeV.}
\label{chi1-psc}
\end{table}

It is found that the condition $\lambda_Y > 0$ can predict the lightest 
glueball scalar should be around 2.0 GeV or so, as can be read off from 
Fig. (\ref{figf}b), while the lightest glueball pseudo scalar should 
be $\eta_5$. The mass splitting between these two glueball states is 
controlled by parameters $v_Y$ and $\lambda_Y$ and is found to be 
around $0.15$ GeV. When compared with the Lattice QCD prediction for 
the glueball bare mass reported in \cite{lattice} where the mass is 
$1.611$ GeV, our result $m_Y=1.73$ GeV is slightly heavier than this 
prediction. When $m_Y=1.611$ GeV is taken, then the predicted 
mass of the lightest glueball is $m_{f_5^0}=2.29$ GeV. 3) The lightest 
scalar $f_1^0(600)$ is found to be $0.27$ GeV or so and is a quarkonia 
dominant state.

In Figure \ref{figf}, we demonstrate the dependence of $\lambda_Y$ and $f^0$ 
masses upon the free parameter $m_Y^2$ with the rest of parameters are given 
in Table (\ref{chi1-prm}). As shown in Fig. (\ref{figf}a), when $m_Y^2$ is 
larger than $3.4$ GeV$^2$, the $\lambda_Y$ becomes negative. Then the 
potential of our model has to confront with the problem of unbounded vacuum 
from below. In the allowed values of $m_Y^2$, the masses of 
$f^0_i\,,i=1,2,3,4$ are almost independent of its value, as demonstrated in 
Fig. (\ref{figf}b). The upper bound of $m_Y^2$ is determined from the 
condition $\Gamma_{f^0_1} > 0.35$ GeV.

To conclude, our model predicts that the isoscalar glueball should be heavier 
than $2.0$ GeV when the pseudoscalar $\eta(1726)$ is the best glueball 
candidates. The lowest isoscalar $f_0(600)$ is found to be quarkonia dominant 
state with a considerable tetraquark component.


\begin{theacknowledgments}
This work is supported by the NSFC under Grant
Nos. 11175251, 11250110058 and 11275213, DFG and NSFC (CRC 110), 
CAS key project KJCX2-EW-N01, CAS 2011Y2JB05, K.C.Wong Education 
Foundation, and Youth Innovation Promotion Association of CAS.
\end{theacknowledgments}



\bibliographystyle{aipproc}   




\end{document}